# Retrieval of aboveground crop nitrogen content with a hybrid machine learning method

Katja Berger[1,*], Jochem Verrelst[2], Jean-Baptiste Féret[3], Tobias Hank[1], Matthias Wocher[1], Wolfram Mauser[1], Gustau Camps-Valls[2]

[1] Department of Geography, Ludwig-Maximilians-Universität Munich, Luisenstr. 37, 80333 Munich, Germany

[2] Image Processing Laboratory (IPL), Parc Científic, Universitat de València, Paterna, València 46980, Spain

[3] TETIS, INRAE, AgroParisTech, CIRAD, CNRS, Université Montpellier, Montpellier, France

*Correspondence: katja.berger@lmu.de

**Abstract:**

Hyperspectral acquisitions have proven to be the most informative Earth observation data source for the estimation of nitrogen (N) content, which is the main limiting nutrient for plant growth and thus agricultural production. In the past, empirical algorithms have been widely employed to retrieve information on this biochemical plant component from canopy reflectance. However, these approaches do not seek for a cause-effect relationship based on physical laws. Moreover, most studies solely relied on the correlation of chlorophyll content with nitrogen, and thus neglected the fact that most N is bound in proteins. Our study presents a hybrid retrieval method using a physically-based approach combined with machine learning regression to estimate crop N content. Within the workflow, the leaf optical properties model PROSPECT-PRO including the newly calibrated specific absorption coefficients (SAC) of proteins, was coupled with the canopy reflectance model 4SAIL to PROSAIL-PRO. The latter was then employed to generate a training database to be used for advanced probabilistic machine learning methods: a standard homoscedastic Gaussian process (GP) and a heteroscedastic GP regression that accounts for signal-to-noise relations. Both GP models have the property of providing confidence intervals for the estimates, which sets them apart from other machine learners. Moreover, a GP-based sequential backward band removal algorithm was employed to analyze the band-specific information content of PROSAIL-PRO simulated spectra for the estimation of aboveground N.

Data from multiple hyperspectral field campaigns, carried out in the framework of the future satellite mission Environmental Mapping and Analysis Program (EnMAP), were exploited for validation. In these campaigns, corn and winter wheat spectra were acquired to simulate spectral EnMAP data. Moreover, destructive N measurements from leaves, stalks and fruits were collected separately to enable plant-organ-specific validation. The results showed that both GP models can provide accurate aboveground N simulations, with slightly better results of the heteroscedastic GP in terms of model testing and against *in situ* N measurements from leaves plus stalks, with root mean square error (RMSE) of 2.1 g/m². However, the inclusion of fruit N content for validation deteriorated the results, which can be explained by the inability of the radiation to penetrate the thick tissues of stalks, corn cobs and wheat ears. GP-based band analysis identified optimal spectral settings with ten bands mainly situated in the shortwave infrared (SWIR) spectral region. Use of well-known protein absorption bands from the literature showed comparative results. Finally, the heteroscedastic GP model was successfully applied on airborne hyperspectral data for N mapping. We conclude that GP algorithms, and in particular the heteroscedastic GP, should be implemented for global agricultural monitoring of aboveground N from future imaging spectroscopy data.

**1. Introduction**

Nitrogen (N) is considered as one of the most important plant macro-nutrients influencing crop growth and quality (Leghari et al., 2016). Therefore, the proper management of N is a pre-requisite for modern agriculture. Optimal crop yield from high quality grain can only be obtained with sufficient N availability and subsequent high uptake of N. Whereas N deficiency leads to a decrease of photosynthetic assimilation and thus crop yield per area, both in terms of quantity and quality (Chlingaryan et al., 2018; Jay et al., 2017), excessive N fertilization causes severe foliar damage (Powell and Lindquist, 1997) and moreover leads to serious environmental problems (Ju et al., 2006; Padilla et al., 2018). In the context of precision farming applications, variable rate N fertilization maps - describing the difference between the actual and optimal N content - support the detection of N deficient areas within the season (Cilia et al., 2014). Hereby, the concept of the nitrogen nutrition index (NNI) was introduced (Lemaire et al., 2008), which can be a valuable tool for the sustainable management of agricultural areas supporting the soil-plant N balance of cropped fields (Cilia et al., 2014).

From the physiological perspective, plants invest large amounts of N into protein and chlorophyll within the leaf cells, with proteins being the major N-containing biochemical constituent (Chapin et al., 1987; Kokaly, 2001). Correlations between leaf nitrogen and leaf chlorophyll content ($C_{ab}$) have been evidenced empirically, which led to $C_{ab}$ content being used as proxy for crop N content. However, moderate relationships between $C_{ab}$ and N content have been reported between species and growth stages across ecosystems, with Pearson correlation coefficients (r) of r = 0.65 $\pm$ 0.15 (Homolová et al., 2013). This may be due to the fact that N is translocated between plant organs during the different development stages of vegetation, independently from chlorophyll. For instance, in case of paddy rice, 80% of total N is absorbed before flowering and bound in vegetative tissues, such as leaves and stalks. Then, during the reproductive phase, N is transported from these vegetative organs to reproductive structures, i.e. the ears, to fill the grain (Ohyama, 2010). Therefore, the relationship between N and chlorophyll content is not maintained after the vegetative growth stage (Gholizadeh et al., 2017). These observations lead to questioning about the limitations of commonly used $C_{ab}$ - N relationships, and motivates the exploration of potentially more robust alternatives for N monitoring.

Leaf protein content ($C_p$) appears as a valid alternative to $C_{ab}$ for N monitoring in the context of future imaging spectroscopy satellite missions (Berger et al., 2020). Imaging spectroscopy sensors with a continuous spectral coverage (hyperspectral) are capable of capturing the subtle spectral signatures related to proteins which are mainly located in the shortwave infrared (SWIR, 1300-2500 nm) region (Curran, 1989; Homolová et al., 2013). Observations from hyperspectral sensors should therefore be preferred over multispectral systems for crop N sensing. In the upcoming years, a number of satellite imaging spectroscopy missions interesting for N monitoring will be launched, as for instance the Environmental Mapping and Analysis Program (EnMAP). EnMAP aims to provide highly accurate hyperspectral imagery of the Earth surface on a regular basis (Guanter et al., 2015) and is planned to be launched in April 2021. The sensor will acquire information over 30km-wide areas in the across-track direction, with 30 m ground sampling distance (GSD). Spectrally, EnMAP will measure around 240 contiguous bands in the optical spectral range using a dual-spectrometer instrument concept with a spectral sampling distance (SSD) between 6.5 nm (420 nm - 1000 nm; VNIR-spectrometer) and 10 nm (900 nm - 2450 nm; SWIR-spectrometer). Relatively high signal-to-noise ratios (SNR) of 400:1 for VNIR and around 180:1 for the SWIR spectrometers are targeted (Guanter et al., 2015). With EnMAP and other missions ahead, such as Italian PRISMA (Loizzo et al., 2019), the Hyperspectral Imager Suite (HISUI) (Matsunaga et al., 2017) or the Italian-Israeli Spaceborne Hyperspectral Applicative Land and Ocean Mission SHALOM (Natale et al., 2013), an era of operational large-scale monitoring systems for agricultural applications will enable the exploitation of subtle biochemical traits such as nitrogen (Hank et al., 2019). Worth to mention in this context is the ESA high-priority candidate mission CHIME, a unique visible to shortwave infrared spectrometer supposed to deliver routine hyperspectral observations for enhanced services, among others for agricultural managements (Nieke and Rast, 2018). For an overview of future missions interesting for N sensing see Berger et al. (2020).

New challenges in data processing and storage arise in view of these new possibilities. Imaging spectroscopy data include correlated and sometimes noisy bands which create statistical problems, also known as curse of dimensionality (CoD) or the Hughes phenomenon (Bellman et al., 1957; Keogh and Mueen, 2017). One of the problems that may arise with the CoD are collinearity effects when apparently independent predictors (bands) are correlated. Hence, in a statistical model the variance of regression parameters is inflated which potentially leads to wrong identification of relevant predictors (Dormann et al., 2013). Further, overfitting is another CoD implying that fluctuations and noise from the training data are learned in high detail during calibration, and flexibility of prediction equations is then in part determined (Altman and Krzywinski, 2018). This frequently happens within statistical models: the established model perfectly applies on internal test data but fails when confronted with unknown values from outside the calibration range (Curran, 1989; Kimes et al., 2000). Moreover, processing of huge amount of data of a hyperspectral image cube can be troublesome, complicating real-time image processing within effective operational hyperspectral imaging solutions (Hogervorst and Schwering, 2011). Training of machine learning methods, for instance, comes with high computational costs when too many bands are involved, as it is usually the case with hyperspectral data. Hence, the reduction of the spectral data space while preserving most essential information could alleviate these drawbacks (Rivera-Caicedo et al., 2017). This is possible with two main methods of dimensionality reduction (DR): (i) feature engineering or transformation, and (ii) feature (band) selection or feature extraction.

In both categories, data are converted into a lower-dimensional feature space, ensuring that the vast majority of the original information is preserved. Feature engineering (i) is usually based on mathematical projections, which attempt to transform the original features into an appropriate feature space. After transformation, the original meaning of the features is usually lost (Wu et al., 2007). The most popular methods within this category are principal component analysis (PCA) (Jolliffe, 2011) partial least squares (PLS) (Wold et al., 1989) or kernel-based alternatives (k-PCA), see (Rivera-Caicedo et al., 2017). With feature selection (ii), a subset of relevant features for model construction is selected without changing the original meaning of the data. In general, a distinction of three different methods can be made: filter, wrapper and embedded modeling (Kohavi and John, 1997; Saeys et al., 2007). A filter approach was for instance applied in the study of Wang et al. (2018), who compared different spectral subsets based on common knowledge or global sensitivity analysis (GSA) for mapping of forest N content, with protein absorption wavelengths from the literature (Curran, 1989; Fourty et al., 1996). The definition of spectral subsets for the retrieval of vegetation traits based upon common knowledge may, however, not provide the highest accuracy. In contrast to filter methods, learning is involved with wrapper (and embedded) models, usually by means of a regression model. In this way, features are ranked according to their meaning for the respective variable, which potentially improves the models' performance. Wrapper techniques iteratively use the outputs of the regression algorithm as selection criteria: in each cycle, the algorithm assures that the selected subset improves the performance of the previous one. In this way, the algorithm explores the feature space and generates various feature subsets (Moghimi et al., 2018; Verrelst et al., 2016b). Since filter methods might fail to select the right subset of bands, if the used criterion deviates from the one used for training the regression algorithm, wrapper methods should be preferred. Different wrapper techniques have been exploited for identification of optimal band settings for the retrieval of biochemical and biophysical variables from proximal hyperspectral sensing at field level and with hyperspectral airborne acquisitions (Atzberger et al., 2013; Berger et al., 2018b; Verrelst et al., 2016b).

Popular methods for N estimation over the last decades were parametric regressions, with narrowband vegetation indices (VI) as the most common approach (Berger et al., 2020). VIs assume an explicit relationship between a spectral observation and the sought biochemical (or biophysical) variable (Verrelst et al., 2019a). Besides chlorophyll-related N estimations using the visible wavelength range (VIS), SWIR absorption features of proteins and N have been exploited by several studies for their capability to predict N, e.g. by Camino et al. (2018), Ferwerda et al. (2005) or Herrmann et al. (2010). Due to limited reliability and transferability of simple parametric regressions across space and time, nonparametric linear methods have been developed, also known as chemometrics (Lavine and Workman, 2013). These include PCA-based regressions (PCR) or partial least square regression (PLSR), the latter being the most widely applied full spectrum chemometric to map vegetation properties and in particular crop N (Berger et al., 2020). Advancing beyond linear transformation techniques, a variety of nonlinear nonparametric methods, also referred to as machine learning regression algorithms, has been developed during the last few decades, e.g. artificial neural networks, random forests or kernel-methods (Camps-Valls et al., 2009). Within the kernel methods, the development of Gaussian processes (GPs) led to substantial accuracy improvements and wide adoption in biophysical and biochemical variable retrievals, as demonstrated by several studies (Camps-Valls et al., 2016; Van Wittenberghe et al., 2014; Verrelst et al., 2012b; Verrelst et al., 2016b). The study of Upreti et al. (2019), for instance, compared different machine learning algorithms and found that GPs were best algorithms in the cross-validation phase for the estimation of classical biochemical and biophysical vegetation traits.

GPs constitute a solid probabilistic framework to develop powerful methods for function approximation problems (Camps-Valls et al., 2019; Camps-Valls et al., 2016). In the context of N estimation, GPs have been used only rarely, for instance by Zhou et al. (2018).

Lastly, physically-based methods using radiative transfer models (RTM) were exploited for N retrieval (Berger et al., 2018c; Li et al., 2018; Li et al., 2019; Wang et al., 2018). In contrast to empirical approaches, RTMs apply physical laws to explain the cause-effect relationships between radiation-photon interactions and plant constituents. By means of these mechanistic models, upscaling from leaf- to canopy level can be accomplished, using for instance the leaf optical properties model PROSPECT-PRO (Feret et al., 2020) coupled with the 1D canopy reflectance model Scattering by Arbitrarily Inclined Leaves, 4SAIL (Verhoef, 1984; Verhoef et al., 2007) to PROSAIL (Jacquemoud et al., 2009). Inversion of the RTM is required to obtain the sought biophysical or biochemical traits. In the past, this has usually been achieved using lookup tables (LUT) or numerical optimization. However, both can be computationally demanding and thus may be less suitable for real time analysis of hyperspectral scenes. Interesting alternatives are currently provided by combining two different techniques, i.e. RTMs and advanced regression methods, such as GPs: simulated data generated by the RTM is used to train the statistical model which is finally applied on real spectral observations. In this way, hybrid methods (Brede et al., 2020; Camps-Valls et al., 2011) use the advantages of both approaches: the underlying physics is regarded by the RTM and at the same time flexibility, scalability and computational speed is provided by the machine learning algorithms (Camps-Valls et al., 2019; Camps-Valls et al., 2016; Verrelst et al., 2015). Hybrid retrieval workflows are therefore very promising for vegetation properties mapping using imaging spectroscopy data and have been implemented in several operational processing chains, mainly in combination with neural networks (Upreti et al., 2019; Verger et al., 2011). Regarding operational production of vegetation traits including N, hybrid methods can compete with simple parametric regression approaches in terms of mapping speed. However, GPs involve high computational costs and memory requirements during training, which exponentially grow with the respective number of training points. Since training in remote sensing problems typically employs a reduced number of samples, both training and test can be done in reasonable times (Rasmussen and Williams, 2006). Despite their strong advantages, hybrid methods have not yet been applied for N retrieval from imaging spectroscopy data.

The present study aims to explore a hybrid retrieval strategy including the coupling of PROSPECT-PRO with 4SAIL, combined with GPs for the retrieval of aboveground N of winter wheat and corn crops. Two spectral sampling strategies are compared for optimal N sensing: a wrapper strategy is implemented to define optimal spectral sampling, and its performance is compared to literature-based protein related spectral wavelengths. By this, we intend to propose a first step towards the generation of operational nitrogen related information products from future imaging spectroscopy data, independent from purely chlorophyll-based relations.

## 2. Materials and Methods

*2.1. Study areas*

2.1.1 Munich-North-Isar (MNI)

An extensive field campaign including field spectroscopy and destructive measurements on crops was carried out in 2017 and 2018 at a test site in the North of Munich, Bavaria, in Southern Germany. During the Munich-North-Isar (MNI) campaign, hyperspectral signatures of the canopy within the 350–2500 nm range were collected from a relatively homogeneous winter wheat (*Triticum aestivum*) and a corn (*Zea mays*) field at almost weekly intervals using the Analytical Spectral Devices Inc. (ASD; Boulder, CO, USA) FieldSpec 3 JR. The sensor provides an effective spectral resolution of 3 nm in the VIS domain from 350-700 nm, and 10 nm from 700-2500 nm, i.e. near infrared (NIR) to SWIR. Reflectance data were collected at nine (3 x 3) elementary sampling units (ESU), which covered a 30 x 30 m area that corresponds to the ground sampling distance of EnMAP as shown in Fig. 1 (upper right). Spectral properties of the EnMAP sensor were simulated using the appropriate spectral response functions (Segl et al., 2012). Due to VNIR/SWIR overlap, some bands were excluded from the analysis, which lead to a final number of 235 bands used for all simulations.

In the same fields and at each date, wheat plants covering an area of 0.25 $m^2$ aboveground were cut, weighed and brought to the lab. In case of corn, three plants were cut and weighed, but only one plant was taken to the lab. Corn-field plant density was obtained by counting plants and rows per meter to relate the final biochemical measurements of the plant samples to unit surface area in [$m^2$]. The Dumas combustion method was applied in the lab, as described by Muñoz-Huerta et al. (2013). There, wheat and corn samples were separated into three different plant organs (leaves, stalks and fruits), weighed in fresh state and oven-dried at 105 °C for 24 h until constant weight was reached. Then, dry weight could be determined (Wocher et al., 2018). The samples were grinded and corresponding N concentration (N%), which refers to the mass of absorbing materials (dry matter) per unit dry mass, given in [mg/g] or [%], was determined. N% of each plant organ was then converted into aboveground N content ($N_{area}$ in g/m²) by multiplying N% with plant organ-specific dry mass per unit ground area. Table 1 indicates the measured ranges, mean values and standard deviations (SD) of aboveground $N_{area}$ of winter wheat and corn for each plant organ as obtained by the MNI campaigns.

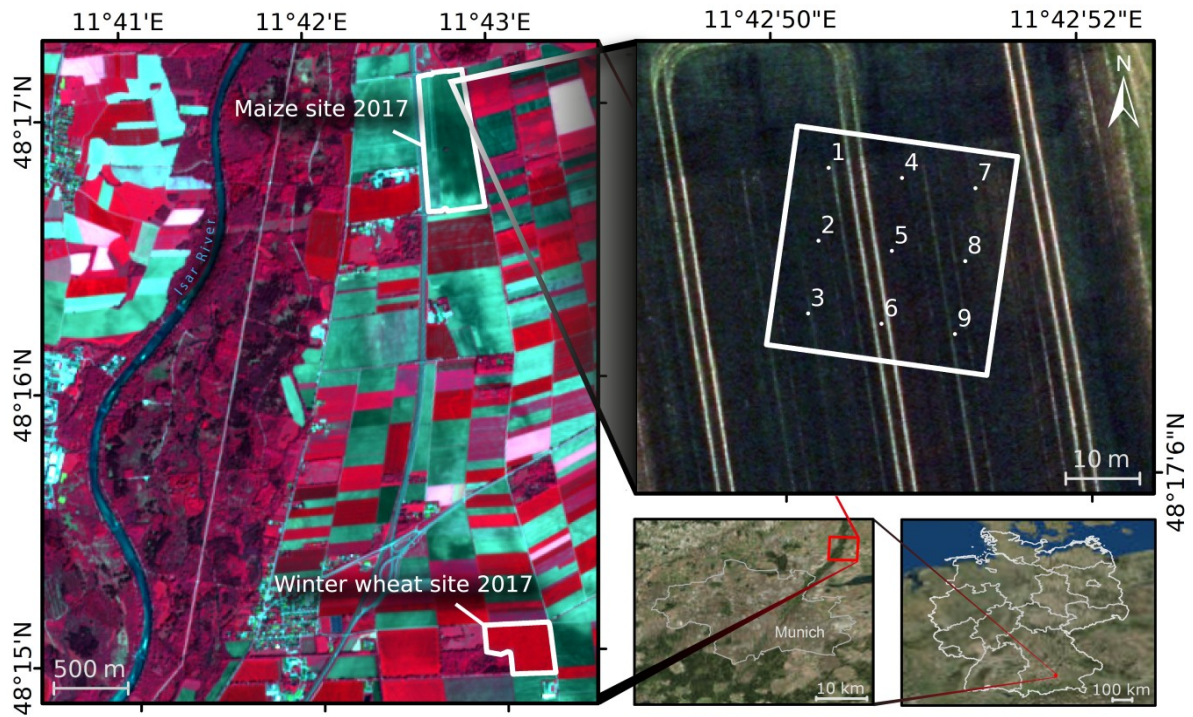

Fig. 1: Munich-North-Isar (MNI) test-sites of corn (maize) and winter wheat: Sentinel-2A false-color-infrared from 17/05/2017 (left), exemplary simulated 30 x 30 m EnMAP-pixel composed of nine ESUs (upper right) and MNI location within Bavaria and Germany (lower right).

Regarding N%, the dilution phenomenon must be considered, meaning that N% decreases with increasing biomass during the growing period. The determination of leaf N% may be important for certain agricultural applications. However, dry matter content is required to enable upscaling to the canopy level (Baret et al., 2007; Kattenborn et al., 2019). Moreover, the biochemical variable $N_{area}$ is highly correlated to the photosynthetic capacity of leaves and thus to carbon fixation (Evans, 1989; Rosati et al., 2000). Therefore, in the current study, we only concentrate on the estimation of aboveground $N_{area}$.

Table 1. Number of measurements (No), range of BBCH growth stages determined according to Meier (2018), *in situ* measured ranges, mean and SD (in brackets) of aboveground $N_{area}$ winter wheat and corn measurements per plant organ at MNI locations.

| Crop / Date | No. | Growth stage [BBCH] | Leaves $N_{area}$ [g/m²] | Stalks $N_{area}$ [g/m²] | Fruits $N_{area}$ [g/m²] |
|---|---|---|---|---|---|
| *Triticum aestivum* 2017, 29/03-06/07 | 9 | 25 - 83 | 2.4 – 13.2 8.4 (3.3) | 0.2 - 8.2 5.5 (2.7) | 0 - 16.4 8.8 (5.1) |
| *Triticum aestivum* 2018, 12/04-13/07 | 7 | 28- 87 | 2.2 - 10.0 6.3 (2.4) | 1.0 - 8.9 4.6 (2.9) | 0 - 25.6 18.3 (5.5) |
| *Zea mays* 2017, 13/06-15/09 | 8 | 30 - 85 | 0.7 - 6.2 4.0 (1.8) | 0.1 - 6.4 2.6 (1.9) | 0 – 13.0 7.3 (5.6) |

Fig. 2 shows the cumulative development of $N_{area}$ for the three organs (leaves, stalks, fruits) of winter wheat (a) and corn (b). The onset of senescence, visible the photographic documentation in Fig. 3 (19/06) for the winter wheat, corresponds to the point in time when $N_{area}$ of the leaves (and stalks) started to decrease and N was translocated to wheat ears. In case of corn, photographic documentation in Fig. 4 indicates the onset of flowering (17/7), which visibly influenced spectral reflectance. N was starting to be translocated to corn cobs from leaves and stalks at this point of time (see also Fig. 2).

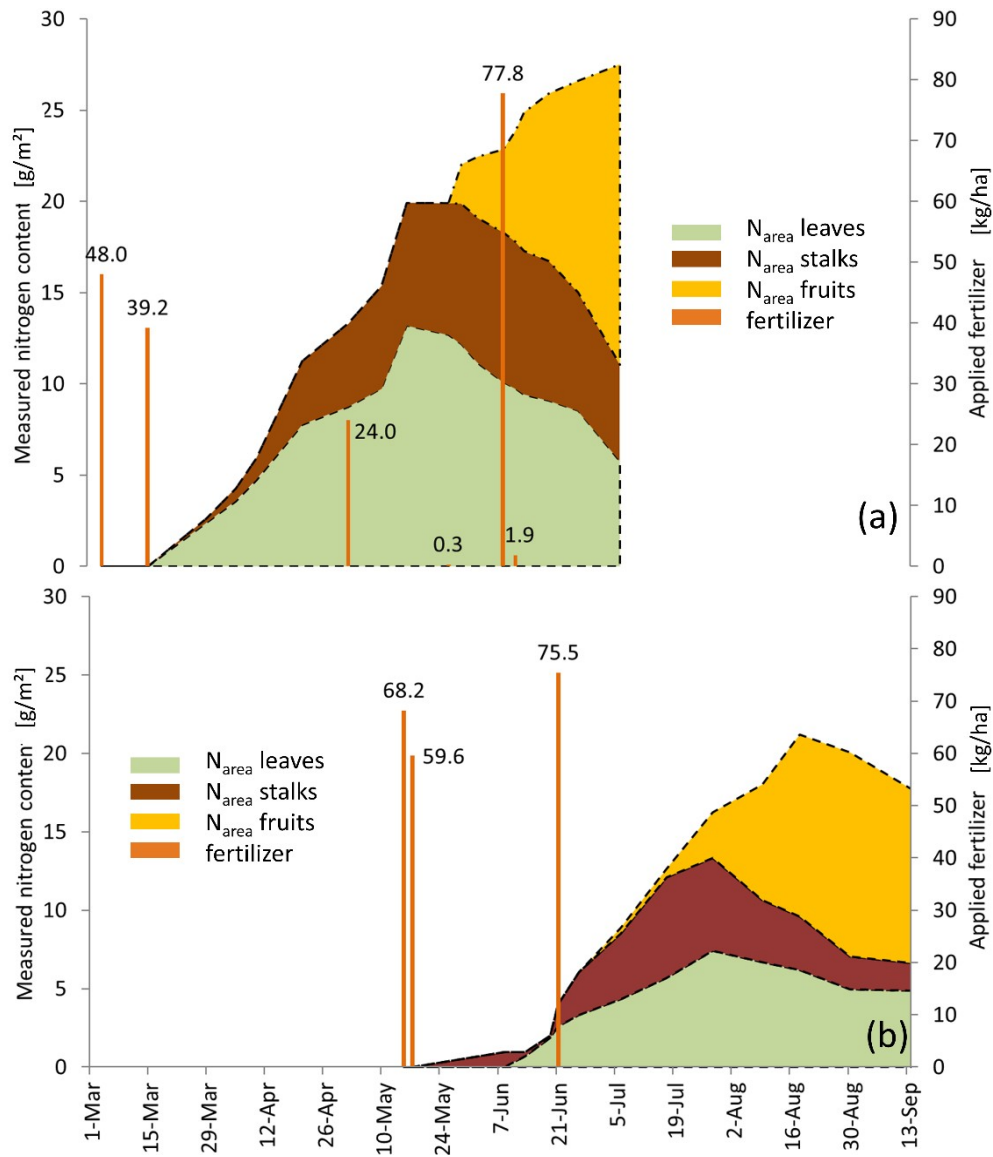

Fig. 2: Development of crop N content of winter wheat (a) and corn (b) with N input via fertilization during growing season in 2017 at the MNI location.

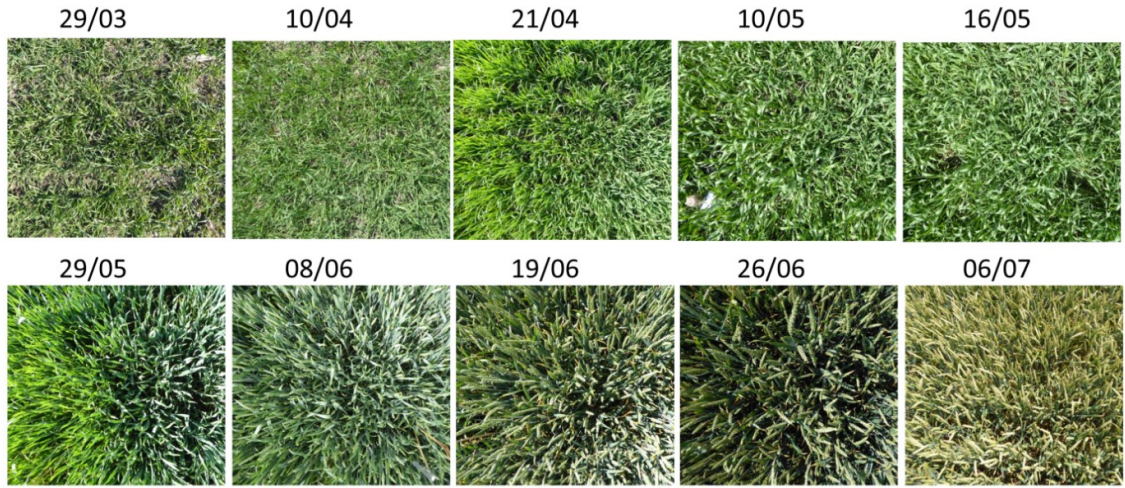

Fig. 3: Photographic documentation of winter wheat growing period from March to July 2017, MNI campaigns, corresponding to biomass sampling dates for N determination. First appearance of senescence at 19/06/2017.

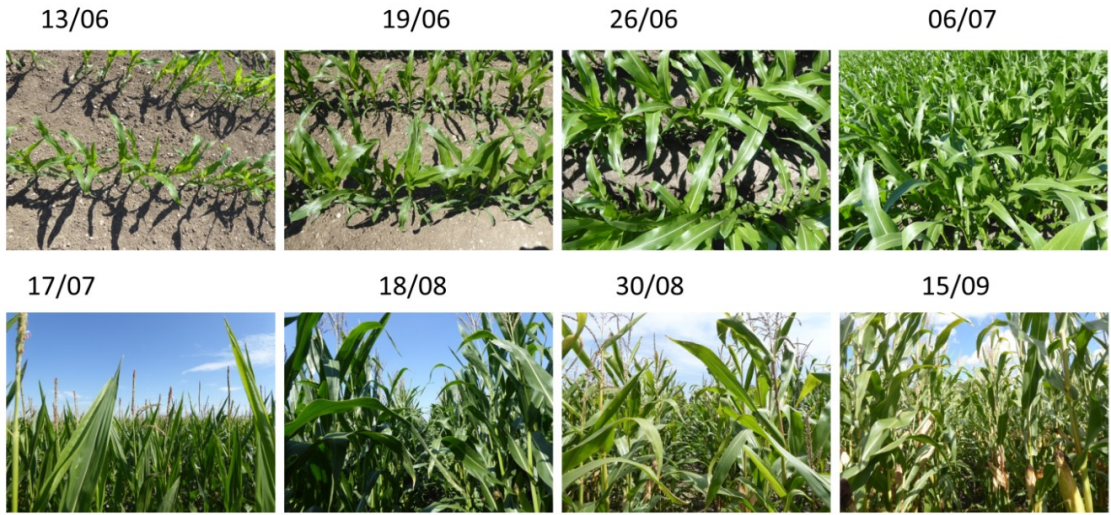

Fig. 4: Photographic documentation of corn growing period from June to September 2017, MNI campaigns, corresponding to biomass sampling dates for N determination. First appearance of flowers at 17/07/2017.

2.1.2 Barrax SPARC data set

ESA organized the SPectra bARrax Campaigns (SPARC) (Moreno et al., 2004) in summer 2003, involving airborne imaging spectroscopy acquisition with the HyMap sensor over Barrax, La Mancha region in Spain (coordinates 30°3"N, 2°6"W). Radiometric and atmospheric corrections were applied (Guanter et al., 2005) in order to produce Top-of-canopy (TOC) reflectance. With a configuration of 126 contiguous spectral bands, spectrally positioned between 438 and 2483 nm, and a spectral bandwidth between 11 and 21 nm, the HyMap sensor is suitable to mimic future spaceborne imaging spectroscopy sensors. HyMAP was resampled to EnMAP spectral configuration using information on full width half maximum (FWHM) bandwidths. The agricultural

Barrax area is situated in a cold semi-arid climate zone. Irrigation is mainly performed with center pivot systems, which results in circular patterns corresponding to fields on imagery data. No *in situ* data of N were collected at this time. However, SPARC was probably one of the most exploited test sites worldwide, and maps of biochemical and biophysical vegetation products were processed by numerous studies, which can be used for comparison purposes (D'Urso et al., 2009; Verrelst et al., 2016b). Here, we selected one flight line from the images acquired on 12$^{th}$ July 2003, in order to illustrate N mapping over an agricultural area.

*2.2 Hybrid retrieval workflow for the estimation of aboveground N*

The design of the hybrid workflow for retrieval of aboveground $N_{area}$ from imaging spectroscopy data, using a RTM database for training GP algorithms, is presented in Fig. 5. The different steps are explained in the following subsections: at first, the RTM PROSAIL-PRO (section 2.2.1) is used to generate a representative LUT (section 2.2.2) for training GP models (section 2.2.3). Selection of optimal bands for $N_{area}$ retrieval is performed with a GP-based band analysis tool (section 2.2.4). Details of training splits and noise levels are explained in section 2.2.5. This retrieval workflow is applicable to top of canopy reflectance (L2A) imaging spectroscopy data including airborne and satellite data. Output corresponds to the predictive mean $\mu*$ of $N_{area}$ for each pixel (*output 1,* Fig. 5) with corresponding retrieval uncertainty, i.e. predictive variance $\sigma^2*$(*output 2,* Fig. 5), which in our case is the SD of the estimates.

2.2.1 Radiative transfer modeling using PROSAIL-PRO

The present study is taking advantage of the leaf model PROSPECT-PRO (Feret et al., 2020) - the latest version of the PROSPECT model (Jacquemoud and Baret, 1990) - capable of separating leaf dry mass per unit leaf area (LMA) into N-based constituents (proteins) and carbon-based constituents (CBC), with the latter including cellulose, lignin, hemicellulose and starch. The calibration of PROSPECT-PRO was based on the principle that proteins and CBC are complementary fractions of total leaf LMA. PROSPECT-PRO was calibrated with fresh and dry leaves. The model has been validated for both kinds of leaves with similar performances for the estimation of protein content. Moreover, the SWIR region was identified as optimal wavelength domain for protein retrieval using common model inversion techniques. Feret et al. (2020) further demonstrated that PROSPECT-PRO is fully compatible with PROSPECT-D (Féret et al., 2017), as it was shown by the indirect estimation of LMA from PROSPECT-PRO inversion obtaining similar retrieval performances when inverting PROSPECT-D. Besides, the author concluded that the PROSPECT-PRO model has a strong potential for the estimation of the carbon/nitrogen ratio (C/N) which could be of interest for diverse agricultural applications in the context of carbon and N sensing using data streams from current and forthcoming satellite imaging spectroscopy missions.

For our purpose, PROSPECT-PRO was coupled with a version of the 4SAIL model (Verhoef et al., 2007) to form 'PROSAIL-PRO' in order to simulate reflectance at the canopy scale as a function of diverse biophysical (e.g. leaf area index, LAI and average leaf inclination angle, ALIA) and biochemical parameters (e.g. $C_{ab}$, $C_p$ or leaf equivalent water thickness, EWT, see also Table 2 for an overview of all input parameters).

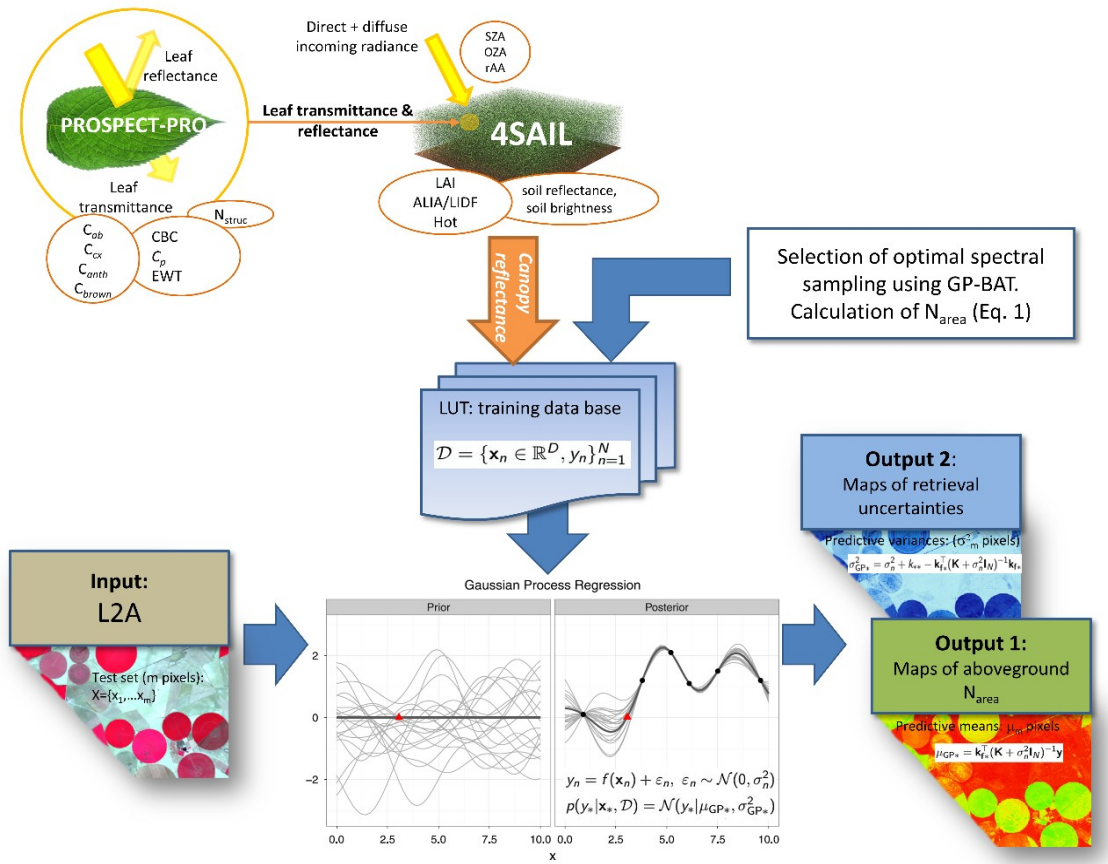

**Fig. 5:** Illustration of inversion scheme using PROSPECT-PRO with 4SAIL models to establish a training data base for the GP models, partly adapted from Berger et al. (2018a), GP figure from Schulz et al. (2018). In the LUT, "*D*" denotes the generated database containing the simulated reflectances $x_n$ and parameter of interest $y$ ($N_{area}$) for *n* pixels. A Gaussian process regression model assumes an additive Gaussian noise model with variance $\sigma_n^2$ and a function *f* that is modeled as a Gaussian process. The inference allows to compute the posterior probability $p(y_*|x_*, D)$ given the new test spectrum $x_*$ and the training dataset *D*. From there we can compute analytically the predictive mean $\mu_*$ and predictive variance (error bars) $\sigma_*^2$ for each pixel, and hence their corresponding spatially explicit maps (Camps-Valls et al., 2016). The input and output maps show our own calculations over the Barrax area in Spain (HyMap, SPARC 2003 campaigns).

2.2.2 Generation of a training data base

The training data base (= lookup table, LUT) was established by randomly generating a total number of 1'000 combinations of all input parameters with PROSAIL-PRO (Table 2). Ranges of the parameters were set according to previous studies and experience of the authors (Berger et al., 2018a; Féret et al., 2017; Verrelst et al., 2016b). Hyperspectral reflectance was simulated corresponding to the EnMAP spectral configurations for the specific parameter space. All parameters were drawn within uniform distributions, except LAI and $C_p$, for which Gaussian distributions were preferred. In this way, a more realistic distribution of $N_{area}$ over the considered growing periods could be simulated; otherwise the distribution of calculated $N_{area}$ would be extremely right-skewed with many low $N_{area}$ values.

Table 2: Parameterization of the simulated data base, i.e. PROSAIL-PRO input parameters with notations, units, and range of parameters to generate EnMAP reflectances, used for the training of the GPs.

| Parameter | Notation (Unit) | Range (min-max) |
|---|---|---|
| *Leaf optical properties (PROSPECT-PRO):* | | |
| Leaf structure parameter | $N_{struc}$, no dimension | 1.0-2.0 |
| Leaf chlorophyll content | $C_{ab}$ [µg/cm$^2$] | 0-80 |
| Leaf water content | EWT [cm] | 0.001-0.02 |
| Leaf carotenoid content | $C_{cx}$ [µg/cm$^2$] | 0-15 |
| Leaf anthocyanin content | $C_{anth}$ [µg/cm$^2$] | 0-2 |
| Leaf protein content | $C_p$ [g/cm$^2$] | 0.001-0.0025 (mean 0.0015, SD: 0.0005) |
| Carbon-based constituents | CBC [g/cm$^2$] | 0.001-0.01 |
| Brown pigment content | $C_{brown}$, no dim. | 0 |
| *Canopy reflectance model (4SAIL):* | | |
| Leaf area index | LAI [m$^2$/m$^2$] | 0–7 (mean: 3, SD: 2) |
| Average leaf inclination angle | ALIA [°] | 30-70 |
| Hot spot parameter | Hot [m/m] | 0.01-0.5 |
| Soil brightness | $\alpha_{soil}$, no dim. | 0-1 |
| Sun zenith angle: | SZA [°] | 30° |
| Observer zenith angle: | OZA [°] | 0° |
| *Calculated ($C_p$, LAI), Eq.1:* | | |
| Aboveground N content | $N_{area}$ [g/m$^2$] | 0.05–29.3 |

Regarding the conversion of simulated protein content into leaf $N_{area}$ content, few previous studies demonstrated that the traditional protein conversion factor of 6.25 is not valid for plant materials due to the existence of non-protein nitrogenous compounds (Milton and Dintzis, 1981; Mosse, 1990; Yeoh and Wee, 1994). Instead, a factor of 4.43 was proposed by Yeoh and Wee (1994) after analysis of 90 plant species. Therefore, as it was also suggested by Wang et al. (2018), we used the conversion factor of 4.43 to calculate $N_{area}$ from modeled $C_p$. Finally, "aboveground $N_{area}$" in [g/m$^2$] was determined from the LUT entries according to equation 1: hereby, leaf $N_{area}$ was calculated from $C_p$ using the above mentioned conversion factor. Multiplication with LAI (and with 10'000) led to the correct ground surface unit by upscaling from leaf level [g/cm²] to canopy level [g/m²]. Finally, simulated N$_{area}$ is comparable to *in situ* values of crop nitrogen, see section 2.1.1.

$$aboveground\ N_{area}\left[\frac{g}{m^2}\right] = \frac{LAI\left[\frac{m^2}{m^2}\right] \cdot Cp\left[\frac{g}{cm^2}\right] * 10'000}{4.43} \quad (1)$$

2.2.3 Machine learning regression models

Gaussian processes (GPs) (Rasmussen and Williams, 2006) are flexible nonparametric models that find functional relationships between input and output variables. These types of machine learning models have excelled in Earth observation (EO) problems in recent years, mainly introduced for model inversion and emulation of complex codes (Camps-Valls et al., 2016; Svendsen et al., 2020). GPs not only provide accurate estimates but also return principled uncertainty estimates for the predictions: due to their solid Bayesian formalism, GPs can include prior physical knowledge about the problem, and allow for a formal treatment of uncertainty quantification and error propagation. Despite the great advantages for modeling, an important challenge in the practical use of GPs in remote sensing problems consists of the fact that data comes with complex nonlinearities, levels and sources of noise, and non-stationarities within covariance functions. The bottleneck of any GP parameterization is the selection of the signal model, and the kernel (covariance) function. In this work, we use two GP models to account for different SNR relations: we used a standard GP model that assumes homoscedastic noise (i.e., a constant noise variance for all observations) and a heteroscedastic GP model that assumes a different noise variance per observation. The standard GPs have excelled in estimating bio-geophysical and biochemical variables from acquired reflectances, ranging from vegetation properties (Mateo-Sanchis et al., 2018; Verrelst et al., 2012a), to dissolved organic matter in water bodies (Ruescas et al., 2018) and atmospheric parameters (Camps-Valls et al., 2012). On the other hand, the current state-of-the-art GP to deal with heteroscedastic noise makes use of a marginalized variational approximation (Lázaro-Gredilla and Titsias, 2011). In both cases, one typically resorts to the use of stationary kernels: several possibilities are available, yet we used the automatic relevance determination (ARD) kernel, that considers different lengthscale hyperparameters per dimension (spectral channel). Both, the noise variance and kernel hyperparameters, are learned either by maximum log-likelihood in the case of the standard GP, see Rasmussen and Williams (2006), or through a variational inference procedure for the heteroscedastic GP, see Lázaro-Gredilla and Titsias (2011) and Lázaro-Gredilla et al. (2014).

2.2.4 Optimal spectral band analysis

In our study, we exploited the property of the ARD covariance in a wrapper strategy using a GP-based band analysis tool, here referred to as "GP-BAT" (Verrelst et al., 2016b). The adapted GP-BAT is based on a sequential backward band removal (SBBR) algorithm and was employed to analyze the band-specific information content of PROSAIL-PRO simulated spectra for the estimation of aboveground $N_{area}$ (see description of LUT, 2.2.2.). GP-BAT includes an iterative backward greedy algorithm, in which the impact of the inputs on the prediction error is evaluated in the context or absence of the other predictors. Essentially, at each iteration, the least significant band is removed (highest $\sigma$), and the algorithms is retrained with the remaining bands. For the dataset, a k-fold cross validation (CV) SBBR procedure was applied. In this way all samples could be used for validation. Goodness-of-fit validation statistics were averaged for the *k* validation subsets, i.e., $R^2_{CV}$, $RMSE_{CV}$, $NRMSE_{CV}$, as well as associated SD and min–max rankings. Based on *k* repetitions, the generated $\sigma_b$ were *k* times ranked. For more details of the GP-BAT procedure refer to Verrelst et al. (2016b).

We further compared the results of the GP-based band analysis with a spectral sampling including protein (nitrogen) absorption bands. These bands were identified and summarized by Curran (1989), Fourty et al. (1996) and Himmelsbach et al. (1988). In Table 3, the EnMAP (central) bands closest to the protein related bands published in the literature are indicated including the absorption mechanisms.

Table 3: Specific absorption bands associated with proteins and nitrogen according to the literature (Curran, 1989; Fourty et al., 1996; Kumar et al., 2001) and selected central EnMAP bands used for training the GP models.

| Protein absorption bands [nm] from literature | EnMAP central waveband [nm] | Absorption mechanism |
|---|---|---|
| 1020 | 1018 | N-H stretch |
| 1510 | 1511 | N-H stretch, $1^{st}$ overtone |
| 1730 | 1731 | C-H stretch |
| 1980 | 1979 | N-H asymmetry |
| 2060 | 2062 | N-H stretch, N=H rotation |
| 2130 | 2133 | N-H stretch |
| 2180 | 2184 | N-H rotation, C-H stretch, C-O stretch, C=O stretch |
| 2240 | 2243 | C-H stretch |
| 2300 | 2300 | C-H rotation, C=O stretch, N-H stretch |

2.2.5 Training splits and noise levels

Random assignment of instances was used for splitting the training and test data for all models. Though there is rather little variation in retrieval accuracy between different training-testing splitting ratios, a clear trend was visible: the smaller the training sample, the larger the RMSE on test data. Instead, the larger the training sample was configured, the lower the test RMSE became. Regarding validation using $N_{area}$ *in situ* data, a model trained over larger training data bases also achieved better results. The final split to be used should be a tradeoff between accurate results of internal learning and performance on real data. The simulated data base was finally split using 70% for training the models and 30% for independent testing. The training partition of the data was used for v-fold cross-validation to select the best hyperparameters for running GP models. This constitutes a strong advantage of using RTM simulations for the establishment of training data bases for models: a large number of all possible situations can be generated including a high degree of variability. Splitting ratio and type of model calibration may play a more important role when using only few *in situ* data for training of the regression models as for instance shown in a previous study by Camps-Valls et al. (2009). However, simulated data were used for training in the present study; hence splitting rate plays a minor role as long as the parameter space is well defined.

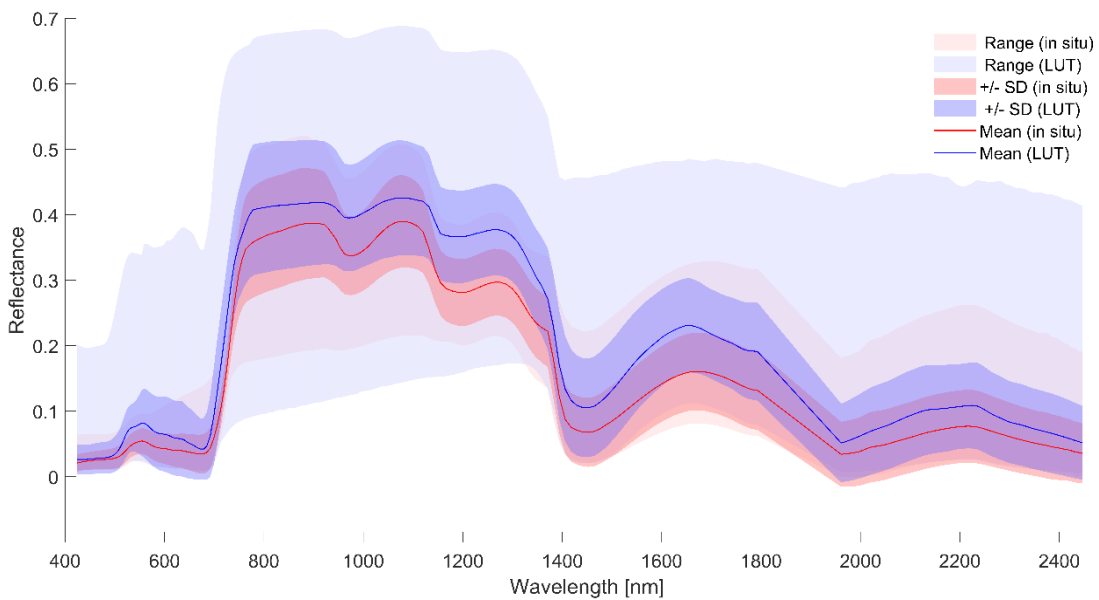

Fig. 6: Illustration of total range of training versus *in situ* reflectances from MNI campaigns, with mean and SD. Training data base (LUT) was simulated with PROSAIL-PRO.

Fig. 6 indicates the ranges, means and standard deviations of the simulated spectra (LUT) used for training the GP algorithm versus *in situ* collected spectral signals. The means of *in situ* and LUT spectra differ slightly, and the two SDs show a small zone of non-overlap. Most importantly, the total range of training spectra comprises the total range of *in situ* collected spectra. This appropriate representation of all possible situations constitutes a prerequisite of a machine learning training data base for successful mapping applications.

Retrieving biophysical and biochemical variables from RTMs is hampered by the presence of high levels of uncertainty and noise (Camps-Valls et al., 2018; Locherer et al., 2015). Consequently, noise has to be considered in the simulated PROSAIL-PRO spectra in order to account for the uncertainty originating from atmospheric conditions, sensor calibration, the radiative transfer model itself (Berger et al., 2018b) or its improper parameterization not completely corresponding to the 'real world' (Danner et al., 2019).

We tested some SD noise levels supplying the same SNR [dB] for all channels to the simulated spectra. Essentially, we varied the SNR, and computed the corresponding SD of the noise, $\sigma_n$, to be added to each channel, which depends on the SD of the channel $\sigma_s$ itself.

## 3. Results

*3.1 Spectral optimization with GP-BAT*

To become familiar with GP band ranking, first the obtained $\sigma_b$ values for a single 3-fold GP model for $N_{area}$ simulated by PROSAIL-PRO, and trained with all EnMAP bands, are illustrated in Fig. 7. The hyperparameter $\sigma_b$ indicates the length-scale per input bands (features), with b ranging from 1 to 235 bands. The inverse of $\sigma_b$ characterizes the relevance of each spectral band; hence, lower values of $\sigma_b$ indicate a higher informative contribution of the respective band b to the training function.

Fig. 7 already suggests the higher importance of the SWIR spectral region, as indicated by lower values of $\sigma_b$. However, there are some drops and peaks, which may be difficult to interpret, and can be due to noise or collinearity. Besides, the $\sigma_b$ values of one GP run may slightly differ when repeated, depending on selected training samples. Therefore, instead of relying on single $\sigma_b$ results for band selection as shown in Fig. 7, a full GP-BAT process was launched and results are demonstrated in Fig. 8. The SBBR procedure removes the least significant band (i.e. highest $\sigma_b$) after each iteration and retrains the algorithm with the remaining bands, as explained in section 2.2.4. Cross-validated coefficient of determination ($R^2_{cv}$) between bands and $N_{area}$ remained stable until 12 bands ($R^2_{cv}$ = 0.98) with SD of 0.006. Below 12 bands it started dropping down to $R^2_{cv}$ = 0.14 (SD=0.033) when using only one band. With one exception, the best performing bands are situated in the SWIR, with center wavelength at 786 nm, 1556 nm, 1568 nm, 1579 nm, 1623 nm, 1656 nm, 1667 nm, 1762 nm, 2124 nm and 2234 nm.

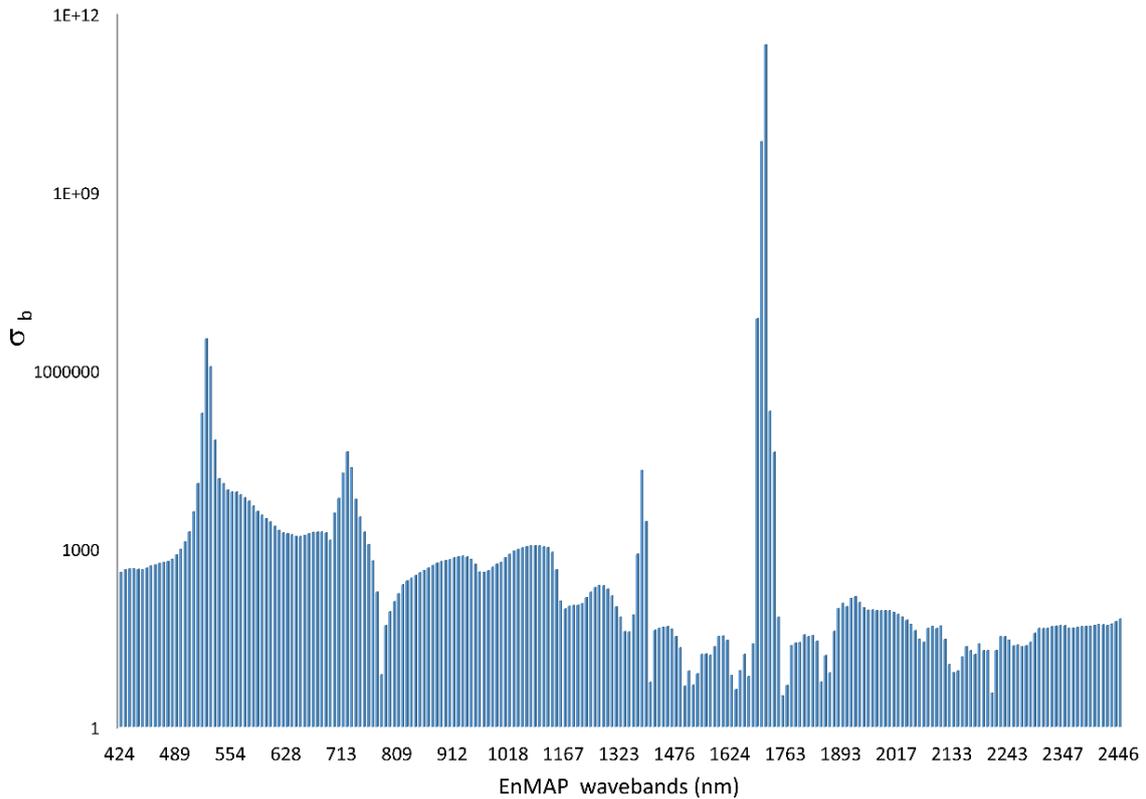

Fig. 7: Band rankings ($\sigma_b$) of all PROSAIL-PRO simulated EnMAP bands for the $N_{area}$ model obtained by a single GP run, plotted in logarithmic scale. Note that the lower the $\sigma_b$ value, the more informative the band is for model development.

Besides, two of the selected 12 bands were located in the atmospheric water vapor absorption region (1834 nm and 1854 nm), which cannot be measured confidentially by hyperspectral sensors; hence they were discarded from further analysis. During the first GP run (Fig. 7), the ten bands already exposed low $\sigma_b$, but were not positioned among the top ten ranks. Instead, the band at 1762 nm with lowest value ($\sigma_b$ =5.0) of the ten selected was ranked at position 5 from 235. The band 786 nm with highest value ($\sigma_b$ =50.5) was ranked at position 83. This emphasizes the importance of the SBBR procedure instead of relying on a sole GP run for band selection. In Fig. 8, the results of the GP-BAT are demonstrated by grouping all selected wavebands into five main spectral classes: VIS, red edge, NIR and two SWIR domains (separated at 1900 nm). The figure can be understood as a histogram, counting the selected bands by GP-BAT from all 235 runs. For this reason, all spectral domains are present, since only one band is removed each time and $N_{area}$ simulation tested with all remaining bands. Quantitatively, most bands were selected from the SWIR 2 region, followed by SWIR 1, NIR, VIS and finally red edge, where the spectral bands were least contributing to $N_{area}$ retrieval. Since these results rely on PROSAIL-PRO simulations and the variable $N_{area}$, which is a combination of LAI and leaf protein content, these results are not surprising but emphasize the importance of the SWIR spectral domain for N retrieval.

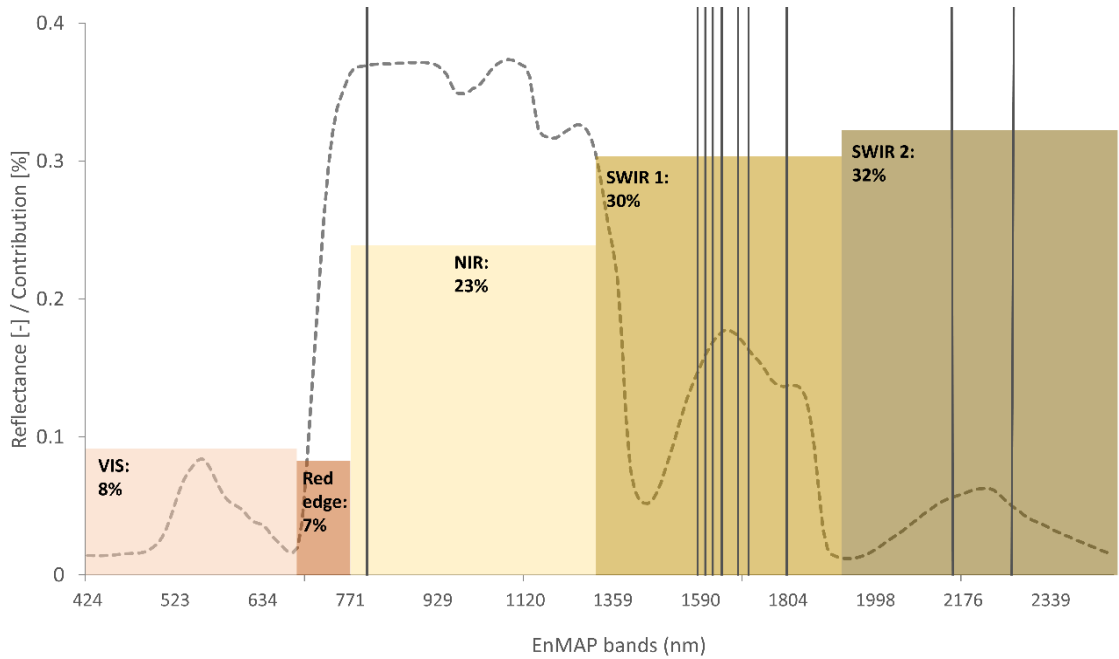

Fig. 8: Contribution of optical domains (VIS: visible, red edge, NIR: near infrared, SWIR 1, 2: shortwave infrared) to the selection of spectral wavebands by GP-BAT with one exemplary PROSAIL-PRO spectrum in the background. Indicated are the ten wavelengths performing best for $N_{area}$ retrieval as identified by GP-BAT regression modelling, with central bands located at 786 nm, 1556 nm, 1568 nm, 1579 nm, 1623 nm, 1656 nm, 1667 nm, 1762 nm, 2124 nm and 2234 nm.

### 3.2 Organ-specific N retrievals

The retrieval of aboveground $N_{area}$ on field measured spectral data using both GP models was tested for the two spectral band settings, including the ten optimal bands found by GP-BAT (section 3.1) and the bands corresponding to protein absorption peaks defined by Curran (1989) and selected by Wang et al. (2018) for N retrieval. Results are demonstrated in Fig. 9, which presents organ-specific aboveground $N_{area}$ retrieved by the heteroscedastic GP model using GP-BAT optimal spectral settings (Fig. 9a-c) and literature-based protein bands (Fig. 9d-f). Fig. 9(a,d) show the accuracy using leaves plus stalks *in situ* $N_{area}$ data for validation with error bars indicating the estimate +/- SD, which corresponds to the associated retrieval uncertainty. Fig. 9(b,e) show validation using only leaves $N_{area}$, and Fig. 9(c,f) illustrate the validation using leaves plus stalks plus fruits, which corresponds to the sum of $N_{area}$ of all plant organs. The retrieval pattern was very similar for both spectral settings, with slightly better results when using GP-BAT selected bands, regarding leaves plus stalks and leaves only. Accuracy achieved by the homoscedastic GP model is similar (results not shown).

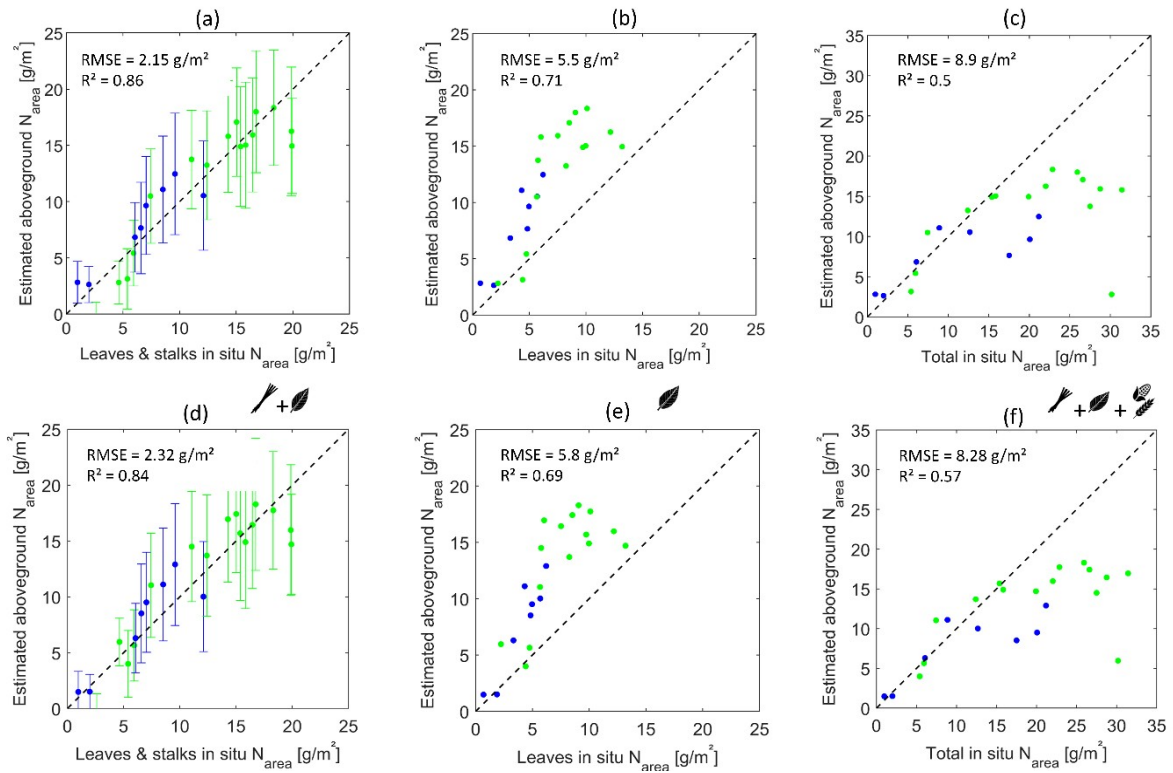

**Fig. 9**: Organ-specific aboveground $N_{area}$ retrieved by the heteroscedastic GP model using GP-BAT optimal spectral settings (a-c) and literature-based protein bands (d-f). (a,d): validation using leaves plus stalks *in situ* $N_{area}$ data with error bars indicating the estimate +/- SD (=retrieval uncertainty), (b,e): validation using only leaves $N_{area}$, and (c,f): validation using leaves plus stalks plus fruits (=sum of $N_{area}$ of all plant organs) as *in situ* reference. Winter wheat is indicated in green, corn in blue dots. Data obtained from MNI campaigns 2017 and 2018.

While the estimation of $N_{area}$ from summed measured leaves and stalks N performed well (Fig.9a,d) strong overestimation occurred when only leaves $N_{area}$ was used as validation data (Fig.9b,e). Instead, strong underestimation occurred preferably for mature growth stages or when fruits were present (Fig.9c,f). Here the inability of radiance to penetrate into the thick plant tissues may play a role (see discussion section 4.2). The scatterplots in Fig. 9(a,d) indicate the uncertainty of the estimates (standard deviation) when using leaves plus stalks $N_{area}$ for validation. Different patterns can be observed between the two GP models: whereas the heteroscedastic GP produces tighter confidence intervals within low-value and higher intervals within high-value regimes, the standard GP provides similar uncertainty levels independent from the estimate (not shown, see discussion section 4.1. and figure in Camps-Valls et al. (2016)). Generally, low uncertainties refer to spectra that were well represented during the learning phase, whereas retrievals with high uncertainties refer to spectral information that deviate from what has been represented during the training stage of the models.

*3.3. Application of the hybrid method to Barrax SPARC03 HyMap imagery*

The trained heteroscedastic GP model using the optimal spectral setting from GP-BAT (section 3.2) was applied to the resampled EnMAP scene from Barrax (original HyMap). Fig. 10 shows the resulting maps of aboveground $N_{area}$ estimates with absolute uncertainties in form of SD. One should keep in mind that SD is related to the magnitude of the mean estimates. Since no flight-parallel *in situ* measured N data were available for the Barrax crops, only a qualitative evaluation can be performed focusing on an exemplary zoomed-in area (see discussion section 4.2). All red-colored areas correspond to fallow or harvested fields or dried out natural vegetation. Probability density plots, which correspond to the continuous version of a histogram with densities, were extracted for one corn and one wheat field to demonstrate the distribution of N within the fields (Fig. 10 lower right). Aboveground $N_{area}$ distribution of these two crops may vary from those of wheat and corn monitored during the whole growing cycles of the MNI campaigns (Fig. 9).

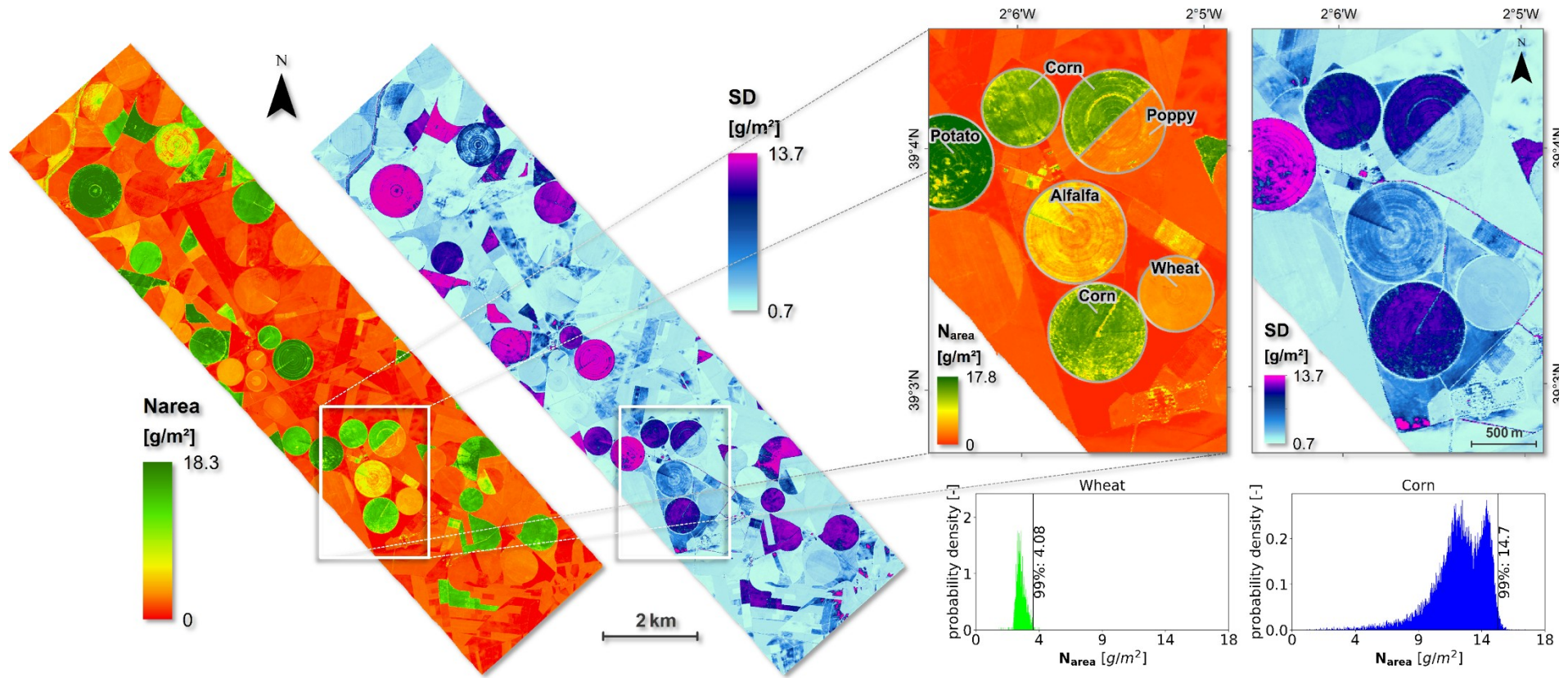

Fig. 10: Resulting maps of aboveground $N_{area}$ generated by the heteroscedastic GP model. Estimates (left) and associated uncertainties (right), expressed as SD around the mean. Simulation from the HyMap scene acquired over Barrax, Spain, during the SPARC campaigns in July 2003. Probability density functions of exemplarily corn and senescent wheat are shown on the lower right. Please note that the $N_{area}$ distribution of both crops may differ from those of wheat and corn monitored during whole growing cycles of the MNI campaigns (Fig. 9). HyMap sensor data was spectrally resampled to EnMAP. Information about SPARC 2003 land use was obtained from D'Urso et al. (2009).

## 4. Discussion

*4.1 Performance of GP models*

Generally, both GP models (standard and heteroscedastic) performed well in learning the nonlinear relationship between reflectance in selected bands according to GP-BAT (as well as literature based) and aboveground $N_{area}$. This confirms the results of other comparative studies as e.g. by Lázaro-Gredilla et al. (2014), who successfully estimated chlorophyll content from hyperspectral data with a heteroscedastic GP model. The latter also (slightly) outperformed the standard GP model in terms of model validation using independent test data and provided most stable results of validation against *in situ* N data. Likewise, the heteroscedastic GP model provided a more differentiated uncertainty pattern of predictive mean and variance in form of tighter confidence intervals within low-value regimes than the homoscedastic GP model. This specific behavior of the heteroscedastic GP uncertainty estimation may be more realistic than provided by the standard (homoscedastic) GP model. For this reasons, the heteroscedastic GP model seems most interesting for N sensing. Generally, this special capability of the GP models to provide variable-associated uncertainties renders them more interesting than other statistical models for global N monitoring using EO data. The provision of retrieval uncertainty information may help the user to better understand how trustful the results are in their context (Weiss et al., 2020). Such information can further indicate the portability of the methods in space and time, and is important when the estimates will be used as input in land surface process models (Camps-Valls et al., 2016). Besides, GPs are endorsed with other attractive properties which are relevant in the field of vegetation properties retrieval, such as the accommodation of different data and noise sources (not shown here).

Both GP models showed a close correspondence between model testing errors and *in situ* validation errors, which is required to avoid under- and overfitting (Kimes et al., 2000). The slightly better results obtained on *in situ* validation (RMSE ≅ 2.2 g/m²) than by model testing (RMSE = 4.5 g/m²) can be explained by the fact that the parameter diversity of the simulated training data base differs from real world: the range of simulated aboveground $N_{area}$ is between 0 and 30 g/m² (see Table 2), whereas the measured range of $N_{area}$ from summed leaves plus stalks is between 1 and 20 g/m² (see Table 1). However, the range of total N content (leaves plus stalks plus fruits) corresponds to the simulated range (measured values from 1 to 30 g/m²). Inability of fruits N detection by hyperspectral reflectance data is discussed in the next section (4.2.). Moreover, some parameter combinations within the LUT may not exist, such as high (green) LAI with low $C_p$. Hence, testing errors may increase with greater proportion of unrealistic combinations of biophysical and biochemical properties in the training dataset. Active learning would be a possible strategy to reduce the dimensionality of a LUT by selecting most informative samples for optimized regression accuracy (Upreti et al., 2019; Verrelst et al., 2016a). We did not exploit this technique in the present study, but employed Gaussian sampling of canopy $N_{area}$ to guarantee realistic sampling of all growth stages (see section 2.2.2). Moreover, LUT size was reduced to 1'000 drawings to provide a compromise between variability of the parameters space and training times of computationally demanding GP models.

*4.2 Organ-specific N retrieval*

Regarding organ-specific retrieval, optimal results could be achieved when *in situ* measurements of leaves plus stalks $N_{area}$ were used for validation (heteroscedastic GP, RMSE = 2.1-2.3 g/m$^2$). When using only leaves $N_{area}$ measurements for validation, strong overestimation occurred (heteroscedastic GP, RMSE = 5.5 - 5.8 g/m²), which is not surprising since stalks, and also the fruits at mature growth stages, contain large parts of total crop N which add to absorption beyond the leaves (Fig. 2). Nevertheless, the tendency to underestimate aboveground $N_{area}$ when fruit $N_{area}$ is included (heteroscedastic GP, RMSE = 8.3 – 9 g/m$^2$) suggests that corn cobs and wheat ears contain more N than it can be detected by a hyperspectral sensor via the reflected signal. In particular, this is the case for mature growth stages, where solar radiation is not able to penetrate the thick tissues of lignified plant material. Penetration depth of incoming solar radiation is also wavelength-dependent. A study analyzing canopy water content (CWC) from the same data set provided similar findings on penetration depth (Wocher et al., 2018). However, this study found validation of CWC retrievals from leaves plus fruits *in situ* data as optimal. This may be due to the different spectral region analyzed: physically-based plant water content retrieval was performed at the 970 nm water absorption depth, embedded in an area of generally high vegetation reflectance and transmittance in the NIR. Nevertheless, our optimal $N_{area}$ retrieval results obtained when using leaves plus stalks *in situ* data for validation, despite the good visibility of wheat ears (see Fig. 3) and (partly) corn cobs (Fig. 4) at the later growth stages, suggests that the results should be interpreted carefully: while not all leaves and stalks $N_{area}$ is 'visible' due to clumping and coverage by the fruits, some of the fruit $N_{area}$ will not be detected due to thick tissues.

These aspects should be considered when interpreting the crop N map from resampled EnMAP image of the Barrax region: retrieved aboveground $N_{area}$ values of cropped fields are in the expected range with absolute maximal values of 18 g/m². In reality though, the total crop N content may be higher due to the tendency of the model to reflect aboveground $N_{area}$ of leaves plus stalks only. Qualitatively, one can compare maps generated by Verrelst et al. (2016b) (see Fig. 7 of their study): in line with the highest green LAI (gLAI), cropped fields also exhibit highest aboveground $N_{area}$ simulations. Realistically detected inter-field variability can be explained with different annual crops and their respective growth stages, see also D'Urso et al. (2009): in the zoomed area of Fig. 10, the two relatively homogeneous bright-green fields (one full and one half) and the lower bright-green field were planted with corn, with gLAI ranging from 3 to 4 m²/m² (D'Urso et al., 2009). Probability density (histogram) of $N_{area}$ is depicted in the lower right of Fig. 10 for one of the corn fields: aboveground $N_{area}$ values from 9-15 g/m² appear realistic for corn being approximately at the same growth stage as the corn at 06/07/2017 (Fig. 4) sampled during the MNI campaigns. The upper left field in the zoomed area was covered with potato, exposing highest aboveground $N_{area}$. This is also reflected by high gLAI values (5-6 m²/m², D'Urso et al. (2009)). Medium $N_{area}$ values of the alfalfa field (see middle of zoomed area, Fig. 10) correspond to gLAI values between 1-3 m²/m². Chlorophyll degradation has already occurred in the wheat fields in senescent growth stages far beyond those analyzed from MNI campaigns (Fig. 3). The obtained maps by Verrelst et al. (2016b) indicate CWC and gLAI values around zero for wheat. Still, the GP models detected quite homogeneously distributed aboveground $N_{area}$ around 2-4 g/m², as shown in the histogram of one exemplary wheat field (see Fig. 10, lower right).

*4.3 Spectral domains*

Influences of confounding factors, such as CWC, on the canopy spectral signal in the analyzed spectral domain may be a potential error source. In our study, we coupled the spectral contributions of leaf protein content with those of LAI, which was identified as the most dominant canopy variable on the total reflectance by a systematic study (Verrelst et al., 2019b). According to the authors, LAI influences the whole spectral range and becomes especially dominant from 1400 nm onwards. Within the 2000–2400 nm spectral region, LAI accounts for 80% to the reflectance and potentially predominates the signal originating from protein absorbance. Our study exploited the spectral information within all identified aboveground $N_{area}$ sensitive bands according to a GP-BAT analysis as well as in relation to protein absorption bands proposed by Curran (1989). We explicitly assume that our approach is able to capture even subtle spectral information of the proteins and thus N (and not only from LAI) as it would never be possible with traditional empirical models (Atzberger et al., 2011). The study of Feret et al. (2020) identified the spectral regions from 2100 nm to 2174/2250 nm as optimal for protein retrieval from leaf optical properties, which corresponds to three of the identified protein absorption bands by Curran (1989) (see also Table 3). The GP-BAT models revealed two bands in this spectral domain as very important (=low $\sigma_b$) for the retrieval of aboveground $N_{area}$ (2124 nm and 2234 nm, see Fig. 8).

The optimal number of bands used for N retrieval should be around nine or ten to avoid under-exploitation of spectral information. Our results showed that this was the case when using less than ten bands provided by GP-BAT, or when using less than the nine bands proposed in the literature (Curran, 1989; Wang et al., 2018). Moreover, when employing a very low number of bands for retrieval (for instance with VI approaches), one could incidentally pick those that are not only sensitive to N (proteins) but also to other biochemicals, such as lignin, cellulose and starch (Curran, 1989; Kumar et al., 2001). Moreover, some bands may be noisy due to the low signal to noise ratio in the SWIR, and thus may not hold sufficient information about proteins. Employing all available spectral bands also provided very low accuracy, probably due to the well-known problem of collinearity. The search of the optimal number of bands for the retrieval of diverse biochemical or biophysical variables from hyperspectral data sets has been discussed in diverse studies. With the same approach (i.e., GP-BAT), Verrelst et al. (2016b), identified six optimal bands for CWC, seven for gLAI and nine for $C_{ab}$ from field-based and airborne data acquisitions. Wang et al. (2008) found an optimal number of 15 bands for LAI retrieval using proximal sensing data. Abdel-Rahman et al. (2013) selected ten wavelengths to estimate sugarcane leaf N% using a machine learning-based ranking approach. Results of these studies are inherently influenced by the applied dimensionality reduction method. The use of feature selection methods for optimal band settings has the advantage of interpretability of the influence of individual features, not given by feature engineering where bands are transformed into incomprehensible components.

*4.4. Advantage and limitations of hybrid methods*

Processing speed of machine learning alleviates the comparably long computational times of mechanistic models, which in turn, contribute with their deductive capability to extrapolate to predictions with behaviors not present in the original data (Baker et al., 2018). Although the proposed combined (hybrid) retrieval workflow will not alleviate the limitations of RTMs, as for instance the ill-posed inverse problem or the constrained model's capability of reproducing the measured (canopy bidirectional) spectral signals (Berger et al. 2018b), it provides the main advantage of providing a comprehensive training data base for the machine learning regression model without the necessity of *in situ* data collection (Verrelst et al. 2019). The LUT can be adapted to the appropriate application by implementing existing knowledge and concepts of experienced users. In fact, the proposed hybrid method can give an indication of crop N which rather corresponds to the amount of N bound in leaves plus stalks. Consequently, the total crop N content may not be obtained in this way. This leads to the conclusion that a more detailed physical (3D) model, embedded within the hybrid scheme, could enhance the retrieval accuracy for total crop N. However, those complex models will also increase the theoretical uncertainties if no prior information about the additional input parameters is available. Complex 3D models would also increase computational times. In order to mitigate this computational burden, intelligent sampling schemes that optimize the LUT size have been proposed, such as active learning procedures (Verrelst et al., 2016a) or adaptive sequential interpolation schemes (Martino et al., 2020; Vicent et al., 2018). In this context, cheap surrogate models of costly complex systems could be established. We strongly suggest to employ such methods within hybrid retrieval workflows to construct small but optimal LUTs for costly RTMs.

Moreover, an atmospheric RTM could be combined with PROSAIL-PRO to improve understanding of the radiative processes occurring in the Earth's atmosphere and to avoid uncertainties from atmospheric corrections (Vicent et al., 2020).

Further, within hybrid retrieval workflows for precision farming applications, mapping of NNI could be introduced. For this, crop- and growth-stage-specific optimal N contents should be known (Cilia et al., 2014).

## 5. Conclusions

Spatiotemporally explicit retrieval methods for high priority biochemical and biophysical vegetation variables such as plant nitrogen are required for numerous Earth system applications. This study was performed in the context of the preparation of the imaging spectroscopy mission EnMAP, which will be an attractive candidate for aboveground crop N monitoring for agricultural regions. When data from EnMAP and in particular future CHIME sensors become available, vegetation properties retrieval workflows will be needed that:

(1) deliver associated per-pixel uncertainties as quality indicators in order to understand if the model can be transferred to other locations and times;
(2) provide high computational speed because repetitive image processing will be needed;
(3) use RTM-based training data since these can be adapted to diverse environmental conditions; and
(4) mitigate the collinearity or band redundancy problems of hyperspectral data using specific feature selection methods.

In our study, we exploited a hybrid retrieval method for aboveground crop N content, composed of two different GP algorithms which were trained over a database simulated with the PROSAIL-PRO model. The heteroscedastic GP model performed best on estimating crop leaves plus stalks *in situ* N data and when optimal spectral settings defined by a GP-based band analysis tool were employed.

GP-BAT has demonstrated to be highly valuable for rapidly identifying relevant band combinations for the estimation of aboveground $N_{area}$. Well-identified band settings not only improve estimation accuracy, but also increase processing speed compared to the use of all available bands from (imaging) spectrometer data.

To test the transferability of our method and to carry out more comprehensive validation, GP models trained for the estimation of aboveground $N_{area}$ should be applied on real imaging spectroscopy (e.g. EnMAP and other) scenes under diverse acquisition geometries, for various locations, crop types and growth stages.

We must emphasize that for proper understanding of the capacity of hyperspectral sensing and algorithms developed to deviate biophysical and biochemical vegetation properties, *in situ* validation data from different plant organs should be available within validation exercises.

From the findings of our study we recommend to combine fast and flexible GP algorithms with RTMs using variable-specific sensitive band settings for the retrieval of the desired vegetation properties. Likewise, our findings demonstrate the possibility to upscale biochemical properties from leaf- to canopy-level using 1D modeling approaches. This constitutes an encouraging step towards the implementation of 3D radiative transfer models, which could provide (more) realistic simulations of the complex interactions between solar radiation in the optical domain with biochemical and structural characteristics of crop canopies. In the near future, it would be of interest to further exploit the combination of 3D RTMs with atmospheric RTMs using intelligent sampling schemes to alleviate computational burden of model generation and GP training pools. These or similar high-level hybrid retrieval workflows uncover great potential towards operational imaging spectroscopy processing schemes for agricultural areas.

**Acknowledgements:**

This research was funded by the German Federal Ministry for Economic Affairs and Energy under the grant code 50EE1623, supported by the Space Agency of the German Aerospace Center (DLR) within the research project EnMAP Scientific Advisory Group Phase III—Developing the EnMAP Managed Vegetation Scientific Processor. Gustau Camps-Valls was supported by the European Research Council (ERC) under the ERC-CoG-2014 SEDAL project (grant agreement 647423). Jochem Verrelst was supported by the European Research Council (ERC) under the ERC-2017-STG SENTIFLEX project (grant agreement 755617) and Ramón y Cajal Contract (Spanish Ministry of Science, Innovation and Universities). J.-B. Féret acknowledges financial support from Agence Nationale de la Recherche (BioCop project—ANR-17-CE32-0001) and TOSCA program grant of the French Space Agency (CNES) (HyperTropik/HyperBIO project). We also thank the two reviewers for their valuable suggestions.